\documentclass[10pt]{article}
\usepackage{ourstyle}
\usepackage{authblk}
\title{Topological and Geometric Reconstruction of Metric Graphs  in $\R^n$}
\author[1,2]{Brittany Terese Fasy}
\author[3]{Rafal Komendarczyk}
\author[3]{Sushovan Majhi}
\author[4]{Carola Wenk}
\affil[1]{School of Computing, Montana State University}
\affil[2]{Department of Mathematical Sciences, Montana State University}
\affil[3]{Department of Mathematics, Tulane University}
\affil[4]{Department of Computer Science, Tulane University}
\date{}

\begin{document}

\twocolumn[{%
    \maketitle
}]

\section{Introduction}
In the last decade, estimation of topological and geometric features
of an unknown underlying space from a finite sample has received an
increasing attention in the field of computational topology and
geometry. For example in \cite{SMALE} the authors provide a
reconstruction guarantee for the topology of an embedded smooth
n-manifold from a finite cover by balls of sufficiently small radius
around a dense enough finite sample. Random sampling is also
considered in \cite{SMALE}, and estimates for the probability of
reconstructing $M$ from a sample are obtained. These estimates imply
that with increasing sample size, the probability of reconstructing
$M$ tends to 1, thus we can recover $M$ almost surely as the sample
size increases to infinity. Also, curve and surface reconstruction algorithms are
discussed in~\cite{Dey:2006:CSR:1196751}

In practice, not all manifolds are smoothly embedded, nor all spaces of interest
are topological manifolds. In \cite{CHAZALSTAB}, the authors show that the
homologies of a compact set $K$ of $\mathbb{R}^n$ can be obtained by considering
only the relevant homological features in the nerve of the $\eps$ radius
balls around a sample that is $\eps$ close to $K$ in the Hausdorff
metric. An upper bound for the required $\eps$ estimate is expressed in
terms of \textit{weak feature size} of $K$ (defined below). The result of~\cite{CHAZALSTAB} works for non-manifolds, but it is limited to spaces that have
a positive weak feature size.

\subsection{Background and related work}
First, we outline a general approach to estimation of topology and geometry of
Euclidean compact sets from the union of balls centered around a finite set of
points densely sampled from the underlying space~$K$. Let $S$ be a sample that
satisfies some density constraints, and let $\eps$ be a radius that depends on
$K$. Then our goal lies to estimate the topology and geometry of~$K$
from~$S^\eps$. Here, $S^\eps$ is the union of Euclidean balls of radius $\eps$
around~$S$.  Roughly speaking, one can expect to capture the topological
features of $K$ if $\eps$ is chosen proportional to the size of the features of~$K$ and proportional to $d_H(S,K)$, the Hausdorff distance between $S$ an $K$.
If $\eps$ is too small or too large, then~$S^\eps$ may fail to capture the
topological features of~$K$.

This suggests the following generic scheme:
\begin{enumerate}
  \item The underlying space $K$ should be a well-behaved space that would allow
    us to choose an appropriate ``feature size'' $\tau$, that would restrict the
    radius $\eps$ not to be too big to capture even the smallest topological
    feature of $K$.
  \item Having $\tau$, for any $\eps<\tau$ one chooses a sample $S$ that
    approximates our underlying space~$K$ very closely, \ie,
    $d_H(S,K)<\eps.$
  \item One then expects to estimate the topology and geometry of~$K$ from
    $S^\eps$.
\end{enumerate}

In \cite{SMALE} the authors show that for a smooth manifold $M$ embedded in
$\mathbb{R}^n$, $\tau_n$ can be chosen to be the maximal radius of the embedded
normal disk bundle of $M$. In \thmref{thm:smale}, the authors show that
$\sqrt{\frac{3}{5}}\tau_n$ can be chosen to be a threshold for $\eps$ to
compute the homologies of $M$ from the union of balls around an
$\frac{\eps}{2}$-dense sample $S$.

\begin{theorem}[Deformation Retraction~\cite{SMALE}]\label{thm:smale}
    Let $M$ be a manifold with injectivity radius $\tau_n$.  Let $\overline{x}$
    be any finite collection of points $x_1,..., x_n\in\mathbb{R}^N$ such that
    it is $\frac{\eps}{2}$ dense in $M$, where
    $\eps<\sqrt{\frac{3}{5}}\tau_n$.  Then, the union $U=\cup_i
    B(x_i,\eps)$ deformation retracts onto $M$.  Consequently, the homology
    of $U$ equals homology of~$M$.
\end{theorem}

We also mention the reconstruction results of \cite{CHAZALSTAB} for compact sets
$K$ in $\mathbb{R}^n$. If $K$ admits a positive weak feature size (wfs) and
$\eps<\frac{1}{4}\wfs$, then a densely sampled set of points can give us
the right topology of $K$.

\begin{theorem}
  Let $K$ and $S$ be two compact sets of~$\mathbb{R}^n$ such that
    $0<\eps<\frac{1}{4}\wfs(K)$ and
  $d_H(K,S)<\eps$. Then,
  $$H_k(K)\backsimeq Image(i),$$ where $H_k$ denotes the $k$-th homology group
  and $i$ is the inclusion of $S^\eps$ in $S^{3\eps}$.
\end{theorem}

\subsection{Summary of results}
The current paper is motivated by the following questions:
\begin{enumerate}
\item \thmref{thm:smale} provides a reconstruction result when
  $\eps<\sqrt{\frac{3}{5}}\tau_n$. Does the result
  fail to hold when $\sqrt{\frac{3}{5}}\tau_n<\eps<\tau_n$?
  
\item The result of \cite{CHAZALSTAB} works for a compact space $K$ having a
  positive \wfs. One can easily find very simple 1-dimensional complexes,
  e.g. trees, that have zero \wfs. These spaces are often of interest in
  applications, for instance in road network reconstruction problems
  \cite{akpw-mca-15}.
  
\item The feature size $\tau$ is defined for smooth manifolds. How can we define
  an appropriate feature size when the space is not a smooth manifold, e.g., an
  embedded simplicial complex.
\end{enumerate}

We answer the first question positively in \thmref{thm:1d}, addressing
the case of smooth curves in $\mathbb{R}^2$. \thmref{thm:1d} shows
that $\eps<\tau_n$ is sufficient for the reconstruction. The second
and third questions are considered in the setting where $K$ is a metric
graph (later denoted by $G$) embedded in $\R^n$. In this
setting, unlike in the manifold case, it is generally not possible to
choose a threshold $\tau$ for the sampling parameter $\eps$ so that
$S^\eps$ has the same homotopy type as $K$, even if $S$ is an
arbitrary dense sample, since $S^\eps$ will contain unnecessary
``small'' features (noise) that are not present in $K$. In order to
address this issue, we propose a different notion of a feature size
that we call \textit{geodesic feature size} (denoted by
$\tau_G$). This new definition of feature size allows us to threshold
the sampling parameter $\eps$, in the case of a metric graph~$G$, and
leads to the reconstruction algorithm shown in \algref{alg:graph}. In
particular, we obtain a simplicial complex $K^\eps$ in $\mathbb{R}^2$,
which is $\eps$-close to $G$ (in the sense of the Hausdorff distance)
and which deformation retracts onto $G$.

\section{Reconstruction Results}\label{sec:results}

\subsection{Smooth Curve Reconstruction}
\begin{theorem}[Smooth Curve Reconstruction]\label{thm:1d}
    Let $M$ be a smooth curve 
in $\mathbb{R}^2$ without boundary and let~$\tau$ be the injectivity radius. Let
$\eps \in (0, \tau]$ and let $S$ be a finite subset of $M$ such that $M\subseteq
  S^\eps$. Then, the medial axis of $S^\eps$ is homeomorphic to $M$.
\end{theorem}

From the result of \cite{lieutier2004any}, which shows that any bounded open
subset of Euclidean space is homotopy equivalent to its medial axis, we conclude
that $S^\eps$ and $M$ are homotopy equivalent.

\begin{remark}
  A deformation retraction constructed in \cite{SMALE}, collapses $S^\eps$
  along the normal lines $M$. The collapse is not well defined if the
  intersection of $S^\eps$ with a normal line has more than one
  connected component. The condition $\eps^2 < \frac{3}{5}\tau_n^2$ (for a
  sample that is $\frac{\eps}{2}$-dense in $M$) guarantees that such
  intersections do not happen and the deformation retraction is
  \mbox{well-defined}.
\end{remark}

\begin{hproof}
    Let $S$ satisfy the assumptions of \thmref{thm:1d}.  For brevity of
    exposition, we assume that~$M$ has one path-connected component.  Then, we
    know that $M$ is, in fact, the image of an injective, smooth map $\gamma
    \colon [0,1] \to \R^2$ with $\gamma(0)=\gamma(1)$. Let us denote the
    $\eps$-tubular
    neighborhood of~$M$ by $M^\eps$.
    
    Without loss of generality, assume that the sample points of $S= \{x_1,
    x_2,..., x_k\}$ are enumerated with increasing preimages: $\gamma^{-1}(x_i)
    < \gamma^{-1}(x_{i+1})$ for all $i \in \{1, 2, \ldots k-1\}$.  Then, these
    samples
    introduce a partition $\{M_i\}_{i=1}^k$ of the manifold, where $M_i =
    \gamma([t_i,t_{i+1}])$ and $M_k = \gamma([t_k,1]) \cup \gamma([0,t_1])$.
  Let $\widehat{M}$ be the piecewise linear curve obtained by connecting $x_i$'s
  in the respective order and let $\widehat{M_i}=\overline{x_i x_{i+1}}$.

    Since $\eps < \tau$, each ball $B_\eps(x_i)$ intersects the tubular
    neighborhood $M^\eps$ at exactly at two points, say at~$u_i$ and $l_i$; see
    \figref{fig:medialAxis}. Let
    $N_i$ denote the normal~$\overline{u_i l_i}$ passing through $x_i$. 
    Notice that these line segments $\overline{u_i l_i}$ do not intersect.  In
    fact, these normal lines partition the tubular neighborhood
    into~$k$ regions, where the $i$-th region, denoted $M_i^\eps$, is the one containing~$M_i$. 

  \begin{figure}[htb]
    \centering \includegraphics[scale=0.75]{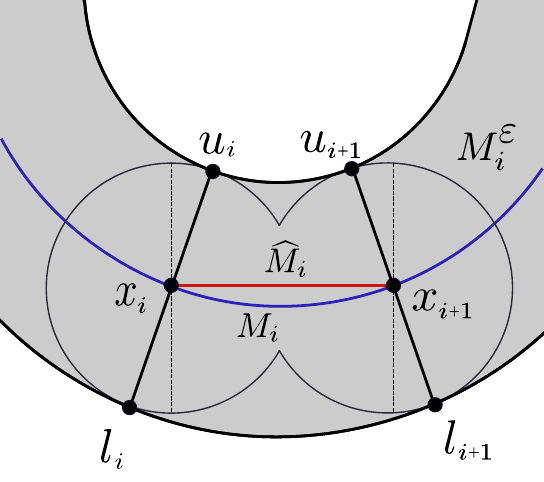}
    \caption{The medial axis and region $M_i^{\eps}$ defined by
        consecutive sample points $x_i$ and $x_{i+1}$.  The blue smooth curve is
        a portion of our manifold~$M$.  The gray region is the tubular
        neighborhood $M^{\eps}$.}
      \label{fig:medialAxis}
  \end{figure}

  We will show that $\widehat{M}$ is homeomorphic to $M$. Observe that
  $\widehat{M}$ is also the medial axis of $S^\eps$.  Restricting our attention
  to $M_i^\eps$, we define a homeomorphism between $\widehat{M_i}$ and $M_i$,
  and extend it globally so that they retain continuity since they agree on
  each~$N_i$ by the pasting lemma.

  We define a homeomorphism $\phi_i:\widehat{M_i}\to M_i$ for each $M_i^\eps$ in
  the following way. If we draw a perpendicular $L$ at any point $z$ on
  $\widehat{M_i}$, we show that~$L$ intersects~$M_i$ at exactly one
  point $y$ and define $\phi_i(x):=y$.  
  As a consequence, $M_i$ is a continuous graph on
  $\widehat{M_i}$, hence a homeomorphism.

  On the contrary, let's assume that there exists a point $x$ on $M_i$ whose
  normal $L$ intersects $M_i$ at at least two points $z_1$ and $z_2$.
  We arrive at a contradiction by showing that there is a point $z$ on $M_i$
  such that the normal $T_z$ at $z$ is parallel to $\widehat{M_i}$.

  Without loss of generality, we assume that $L$ cuts the manifold at both $z_1$
  and $z_2$.  
    Note that $z_1$ and $z_2$ are points on the manifold and tangents
  $T_{z_1}$ and $T_{z_2}$ are not parallel to $L$.  
  By continuity of the tangents of $M$, we conclude that there exists a point
  $z$ on $M_i$ such that~$T_z$ is parallel to $L$. Consequently, the normal
  $N_z$ at~$z$ is parallel to~$\widehat{M_i}$.  
  \begin{figure}[htb]
    \centering
    \includegraphics[scale=0.8]{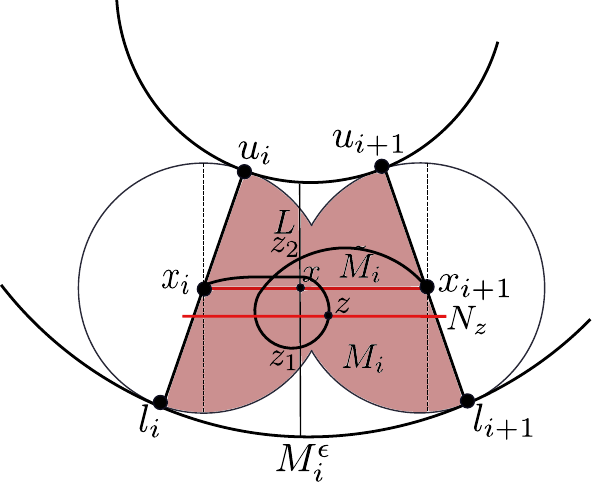}
    \caption{Contradiction in the proof of \thmref{thm:1d}.}\label{fig:contradiction}
  \end{figure}
  
  Now, we arrive at a contradiction in either of the following cases; see
  \figref{thm:1d} for an illustration.

  Case 1:
  If $||x_{i+1}-x_i||\leq\eps$, then the $\eps$-radius normal, $N_z\cap
  M^\eps$, at $z$ intersects either $N_i$ or $N_{i+1}$. This contradicts
  the fact that $\tau$ is the injectivity radius.

  Case 2:
  If $||x_{i+1}-x_i||>\eps$, then the $\eps$-radius normal, $N_z\cap
  M^\eps$, at $z$ lies completely in the interior of~$M_i^\eps$, which 
  is a contradiction because the boundary of each $\eps$-radius normal
  lies on the boundary of the tubular neighborhood $M^\eps$ of the
  manifold.
  
  Therefore, the function $\phi_i$ is a well-defined, invertible, continuous map on a compact
  domain, hence a~homeomorphism.
  
  Since the $\phi_i$'s agree on the boundary of each $\widehat{M_i}$, we glue them
  to obtain a global homeomorphism $\phi:\widehat{M}\to M$. This completes the
  proof.
\end{hproof}


\subsection{Metric Graph Reconstruction}
A weighted graph $G=(V,E)$ is said to be a \emph{metric graph} if the
edge-weights are all positive.  Then, we can interpret the weights as lengths,
and thus each point $e \in E$ has a well-defined distance to the endpoints of
$E$.  We define the length of a given continuous path in $G$ to be the total
length of all edges and partial edges in the path.  Then, the distance function
$d_G \colon G \times G \to \R$ is defined to be the length of the shortest path
connecting two points; in words,~$d_G$ is the geodesic distance in $G$.  Metric
graphs were first introduced in~\cite{kuchment2004quantum}, and have recently
been studied in~\cite{aanjaneya2012metric,gasparovic2017complete}. 

Below we list our assumptions about the underlying graph $G$ that we aim to
reconstruct.

\begin{assumption}\label{ass:graph}
  $G$ is an embedded metric graph with straight line edges.
  $V=\{v_1, v_2,..., v_n\}$ is the vertex set and $E=\{e_1, e_2,..., e_m\}$ is the edge
  set.
\end{assumption}

\begin{assumption}\label{ass:length}
  The length of the smallest edge of $G$ is $l$.
\end{assumption}


\begin{definition}[Nerve of a Cover]
  Suppose $\mathcal{U} = \{U_\alpha\}_{\alpha\in\Lambda}$ is a cover of a
  topological space $X$. We take $\Lambda$ to be the vertex set and form an
  abstract simplicial complex $\mathcal{K}$ in the following way: if a $k$-way
  intersection $U_{\alpha_1}\cap U_{\alpha_2}\cap...\cap U_{\alpha_k}$ is
  non-empty, then $\{\alpha_1, \alpha_2,..., \alpha_k\}\in\mathcal{K}$.

  $\mathcal{K}$ is then called the \emph{nerve} of the cover $\mathcal{U}$
  and is denoted by $\mathcal{N}(\mathcal{U})$.
\end{definition}

\begin{lemma}[Nerve Lemma~\cite{Alexandroff1928}]\label{thm:nerve}
  Suppose\\ $\mathcal{U} = \{U_\alpha\}_{\alpha\in\Lambda}$ is a ``good'' covering
  of $X$, i.e., every $U_\alpha\in\mathcal{U}$ is contractible along with all
  non-empty finite intersections of elements of $\mathcal{U}$. For such a good
  covering $X$ has the same homotopy type as $\mathcal{N}(\mathcal{U})$.
\end{lemma}

We now propose our feature size that we call \\ Geodesic Feature Size (\gfs).
\begin{definition}[Geodesic Feature Size]\label{def:gfs}
  Let $G$ be an embedded metric graph. We define the Geodesic Feature Size
  (\gfs) $\boldsymbol{\tau_G}$ of $G$ to be the supremum of all $r>0$ having the
  following property: for any $x$, $y\in G$, if $||x-y||<2r$ then $d_G(x,y)<l$,
  where $l$ is the length of the smallest edge of $G$.
\end{definition}

To motivate the above definition of \gfs, we take a finite sample $S = \{x_1,
x_2, \ldots, x_k \}$ from $G$.  Let $\{\B_\eps(x)\}_{x \in S}$ be a cover of $G$
and let $\mathcal{K}_1=\mathcal{N}(S,\eps)$ be its nerve, where $\B_\eps(x)$ is
the Euclidean $\eps$-ball centered at $x$.  An edge $e$, between two vertices $x_i$ and
$x_j$ in~$\mathcal{K}_1$, is called \emph{transverse} if $x_i$ and $x_j$ belong
to two different edges of $G$. If $\eps<\tau_G$ and $e$ is transverse, then the
geodesic distance $d_G(x_i,x_j)<l$ and the geodesic on $G$ is unique. This
implies that there is at most one vertex $v$ of~$G$ lying on this geodesic. We
call this geodesic the geodesic shadow of the edge $e$.  The threshold $\tau_G$
for $\eps$ forces any transverse edge $e$ to be within $\B_{\xi\epsilon}(v)$
around that vertex $v$, where $\xi=\max\frac{1}{\sin{(\alpha/2)}}$, the maximum
is taken over all acute angles $\alpha$ between any pair of edges of $G$. The
idea behind this definition of $\gfs$ comes from our goal to estimate the
diameter of non-trivial 1-cycles of $\mathcal{K}_1$ that are not present in $G$.
These \emph{noisy} one-cycles in~$\mathcal{K}_1$ are formed by some of the
transverse edges. We also mention here without a proof that $\gfs$ is positive
for the type of metric graphs we are considering. In fact, we can show that
$0<\tau_G\leq l/2$, where~$l$ is the length of the shortest edge in $G$.

We now state our main reconstruction theorem for embedded metric graphs. This
theorem proves the correctness of \algref{alg:graph} for computing the
1-dimensional Betti number of $G$.

\begin{theorem}\label{thm:graph}
  Let $\eps<\frac{1}{\xi}\gfs(G)$ and $S$ be a finite sample from $G$ such that
  $S^\eps\supseteq G$. Then, $H_1(G)= i_*(H_1(S^\eps))$, where $i$ is the
  inclusion map from $S^\eps\to S^{\xi\eps}$, $H_1(\cdot)$ denotes the first
  homology group in $\Z$ coefficients and $\xi$ is as defined above.
\end{theorem}

\begin{hproof}
  Let $\mathcal{K}_1= \mathcal{N}(S,\eps)$ and $\mathcal{K}_2=
  \mathcal{N}(S,\xi\eps)$.  As $\xi>1$, it follows that $\eps<\gfs(G)$. An
  application of the nerve lemma implies that there is an injective homomorphism
  $\phi$ from $H_1(G)$ to $H_1(\mathcal{K}_1)$.  In other words, $\mathcal{K}_1$
  contains all the non-trivial 1-cycles of $G$. Similarly, we can show that
  there also exists an injective homomorphism from $H_1(G)$ to
  $H_1(\mathcal{K}_2)$. We then consider the induced homomorphism
  $i_*:H_1(K_1)\to H_1(K_2)$, where $i:\mathcal{K}_1\to\mathcal{K}_2$. Finally,
  we show that $H_1(\mathcal{K}_1)=\phi(H_1(G))\oplus Ker~i_*$. Therefore,
  $H_1(G)\simeq i_*(H_1(S^\eps))$.
\end{hproof}

We believe that it should be possible to find a simplicial complex with the same
homotopy type as~$G$. We formulate this stronger result as follows:

\begin{conjecture}
  Let $\eps<\frac{1}{2(2+\xi)}\gfs(G)$ and $S$ be a finite sample from $G$ such
  that each edge of $G$ can be covered by the union of $\eps$-balls centered
  at the sample points on the same edge. Then the Vietoris-Rips complex
  $VR(S,d_\eps, 2(1+\xi)\eps)$, computed on $S$ w.r.t. the geodesic metric on
  the 1-skeleton of $\mathcal{K}_1$ at a scale of $2(1+\xi)$, has the same homotopy
  type as $G$.
\end{conjecture}

\begin{remark}
  The idea of collapsing the ``small'' 1-cycles in~\algref{alg:graph} motivates
  us to add a full simplex around each vertex whenever a subset of $S$ has a
  diameter smaller than the estimated scale. That is precisely what the
  Vietoris-Rips complex does on a finite metric space.
\end{remark}

\subsection{Discussion}
To further extend our result, we also consider a probabilistic reconstruction,
as considered by the authors of~\cite{SMALE}. Given a $(1-\delta)$ chance of
correct reconstruction, one can find the smallest sample size to guarantee the
given chance of recovery. Also, we can extend our definition of \gfs\ to
metric graphs and obtain similar reconstruction
results. Lastly, we also consider the reconstruction question when samples that
are drawn not exactly from our underlying space, but from a close vicinity of
it.

\begin{algorithm}[t]\label{alg:graph}
  \KwData{The finite sample $S$ from the unknown metric graph $G$ and $\eps>0$}
  \KwResult{one dimensional Betti number of $G$}
  Compute $\mathcal{K}_1$
  Compute $\mathcal{K}_2$
  
  \For{non-trivial 1-cycle $\eta$ in $\mathcal{K}_1$} { 
      \If{$\eta$ is trivial in $\mathcal{K}_2$} 
      {Collapse $\eta$ in
      $\mathcal{K}_1$ }
    
  } 
  Compute remaining non-trivial cycles in~$\mathcal{K}_1$\;
  \caption{Metric graph reconstruction from a finite sample.}
\end{algorithm}

\section{Acknowledgments}
The first, third, and fourth authors would like to acknowledge the
generous support of the National Science Foundation under grants
CCF-1618469 and CCF-1618605.

\footnotesize{ 
\bibliographystyle{acm} 
\bibliography{../../thesis} 
}

\end{document}